\documentclass[aps,prx,twocolumn,amsmath,amssymb,superscriptaddress,floatfix]{revtex4-2}

\usepackage{multirow}
\usepackage{xcolor}
\usepackage{bm}
\usepackage{amsfonts,amssymb,amsmath}
\usepackage{graphicx,dcolumn,bm,color,braket,slashed}
\usepackage{times} 
\usepackage{comment}
\usepackage{array}
\usepackage{textcomp}

\begin{document}

\title{Flat bands in Network Superstructures of Atomic Chains}

\author{Donghyeok Heo}
\affiliation{Department of Physics, Ajou University, Suwon 16499, Korea}

\author{Jun Seop Lee}
\affiliation{Department of Physics, Ajou University, Suwon 16499, Korea}

\author{Anwei Zhang}
\email{zawcuhk@gmail.com}
\affiliation{Department of Physics, Ajou University, Suwon 16499, Korea}

\author{Jun-Won Rhim}
\email{jwrhim@ajou.ac.kr}
\affiliation{Department of Physics, Ajou University, Suwon 16499, Korea}
\affiliation{Research Center for Novel Epitaxial Quantum Architectures, Department of Physics, Seoul National University, Seoul, 08826, Korea}

\begin{abstract}
We investigate the origin of the ubiquitous existence of flat bands in the network superstructures of atomic chains, where one-dimensional(1D) atomic chains array periodically.
While there can be many ways to connect those chains, we consider two representative ways of linking them, the dot-type and triangle-type links.
Then, we construct a variety of superstructures, such as the square, rectangular, and honeycomb network superstructures with dot-type links and the honeycomb superstructure with triangle-type links.
These links provide the wavefunctions with an opportunity to have destructive interference, which stabilizes the compact localized state(CLS).
The CLS is a localized eigenstate whose amplitudes are finite only inside a finite region and guarantees the existence of a flat band.
In the network superstructures, there exist multiple flat bands proportional to the number of atoms of each chain, and the corresponding eigenenergies can be found from the stability condition of the compact localized state.
Finally, we demonstrate that the finite bandwidth of the nearly flat bands of the network superstructures arising from the next-nearest-neighbor hopping processes can be suppressed by increasing the length of the chains consisting of the superstructures.
\end{abstract}

\maketitle

\section{Introduction}
A flat band denotes a band with a zero group velocity over the whole Brillouin zone~\cite{leykam2018artificial,rhim2021singular}.
When the flat band becomes slightly dispersive, which is the case in real experiments, we call it a nearly flat band~\cite{wang2011nearly}.
The flat band systems have received great attention because of their intriguing many-body and geometric aspects.
When the Coulomb interaction between electrons is introduced, the flat bands host unconventional superconductivity~\cite{volovik1994fermi,cao2018unconventional,liu2021spectroscopy,balents2020superconductivity,peri2021fragile,yudin2014fermi,volovik2018graphite,aoki2020theoretical,kononov2021superconductivity}, ferromagnetism~\cite{mielke1993ferromagnetism,tasaki1998nagaoka,mielke1999stability,hase2018possibility,you2019flat,saito2021hofstadter,sharpe2019emergent}, Wigner crystal~\cite{wu2007flat,chen2018ferromagnetism,jaworowski2018wigner,rhim2015analytical}, and fractional Chern insulator~\cite{tang2011high,sun2011nearly,neupert2011fractional,sheng2011fractional,regnault2011fractional,wang2011nearly,weeks2012flat,trescher2012flat,yang2012topological,liu2012fractional,bergholtz2013topological}.
The quantum distance, one of the geometric quantities of the Bloch wavefunction, plays an important role in the anomalous Landau levels~\cite{rhim2020quantum,hwang2021geometric}, a new kind of bulk-interface correspondence~\cite{oh2022bulk}, and appearance of the topological non-contractible loop states in flat band systems~\cite{ma2020direct}.
Moreover, it was revealed that the quantum metric~\cite{hwang2021wave} is the key quantity in the physics of the superfluidity~\cite{peotta2015superfluidity,julku2016geometric} and orbital magnetic susceptibility~\cite{raoux2015orbital,piechon2016geometric}.

Despite the numerous intriguing properties of the flat band, it has become a popular research subject only recently since the experimental realization of the nearly flat bands in the twisted bilayer graphene at the magic angle~\cite{cao2018unconventional}.
In addition to this, many artificial flat band systems have been examined~\cite{ma2020direct,guzman2014experimental,vicencio2015observation,mukherjee2015observation,xia2018unconventional,leykam2018perspective,xie2021fractal,song2022topological}, lots of nearly flat band materials such as CoSn~\cite{kang2020topological,liu2020orbital} and FeSn~\cite{kang2020dirac} have been synthesized~\cite{yin2019negative,ye2018massive,lin2018flatbands,yang2019evidence,wang2020experimental}, and many candidate materials have been proposed theoretically recently~\cite{hase2019flat,skurativska2021flat,de2022vacancy,xiao2021flat,kennes2020one,yamada2016first,regnault2022catalogue}.
We focus on the frequent appearance of flat bands in lattice structures with a large-size unit cell such as cyclic-graphyne, cyclic-graphdiyne, and honeycomb network in the nearly commensurate charge-density-wave phase of 1T-TaS$_2$~\cite{you2019flat,lee2020stable}.
Since these lattices are in the shape of a periodic network of finite-size 1D chains, they are called \textit{network superstructures}.
The common feature of network superstructures is that they host extremely flat bands at multiple energy levels.
Although the existence of the flat bands in several network superstructures such as those mentioned above was reported already~\cite{you2019flat,lee2020stable}, the general understanding of why the flat bands are so ubiquitous in this class of systems and why there are multiple numbers of flat bands are developed is not studied yet.

In this paper, we understand the existence of flat bands in the network superstructures from the perspective of the special localized eigenmode of the flat band, so-called the compact localized state(CLS)~\cite{rhim2019classification,hwang2021flat,hwang2021general}.
The CLS is characterized by the fact that it has nonzero amplitude only inside a finite region in real space while exactly zero outside it. 
It was rigorously shown that one can always construct a CLS by a linear combination of the Bloch wave functions of a flat band~\cite{rhim2019classification}.
If the Bloch wave function of the flat band does not have any singularity in momentum space, one can find $N$ number of linearly independent CLSs to span the flat band completely, where $N$ is the number of unit cells in the system.
However, when the Bloch wave function becomes discontinuous due to a band-crossing with another band, the CLSs cannot form a complete set, and they are always linearly dependent.
In this case, some non-compact eigenstates independent of the CLSs are required to exist to form a complete set of eigenfunctions spanning the flat band.
Such non-compact states are usually found as non-contractible loop states(NLSs), which are extended along one spatial direction while compactly localized along another direction.
NLSs show topological features in real space because they cannot be cut by adding CLSs and exhibit a winding feature over the whole system under the periodic boundary condition.
This kind of flat band is called a \textit{singular flat band}.

In the network superstructures, the CLS can be stabilized due to the destructive interference offered by the special lattice structure around the linking parts between 1D chains.
We consider two representative types of the linking structure, the dot-type, and triangle-type.
We construct various network superstructures by connecting the 1D chains in diverse ways and then show that they host many flat bands.
The number of flat bands equals to the number of independent CLSs.
While the flat band becomes a nearly flat band when some long-range hopping processes are introduced, we show that the bandwidth of the nearly flat band can be suppressed by increasing the length of the 1D chains.

\begin{figure*}[ht]
    \centering
    \includegraphics[width=2.0\columnwidth]{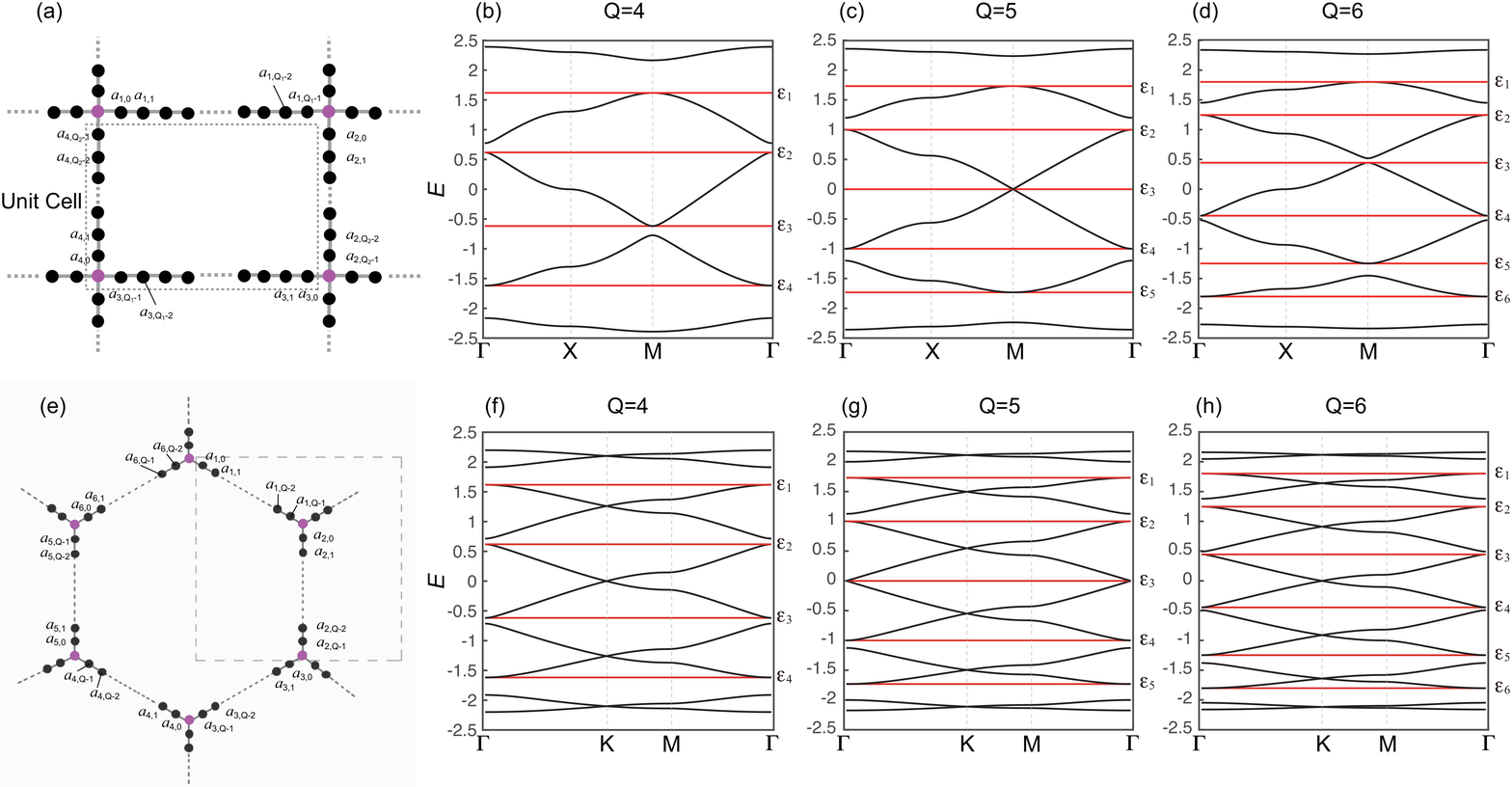}
    \caption{(a) and (e) plot the rectangular and honeycomb network superstructures with dot-type links respectively. When $Q_1=Q_2$, the rectangular network superstructure becomes the square network superstructure. Dot-type links are represented by purple color. $a_{m,n}$ reads the amplitude of the CLS. In the network superstructures with dot-type links, the CLS's amplitude at the dot-type link is always zero. In the unit cell of the rectangular network, we have $Q_1+Q_2+1$ sites, while we have $3Q+2$ sites in the honeycomb case. From (b) to (d) and (f) to (h), we plot band structures of the square and honeycomb network superstructures with different sizes $Q$. Here, red lines represent the flat bands, whose energies are denoted by $\varepsilon_n$. Here, only the nearest-neighbor hopping processes are considered.}
    \label{fig:dotted_basic}
\end{figure*}

\section{Network superstructures hosting flat bands}
Since the existence of the flat band is equivalent to the presence of the CLS, we can obtain a flat band model by designing a lattice structure stabilizing the CLS.
While the CLS has zero amplitudes outside a finite region, it should maintain the same shape after the hopping processes to be an eigenmode.
To this end, destructive interference is necessary to avoid any dissipation of the amplitudes of the CLS after the hopping processes.
As an example, the kagome lattice with the nearest-neighbor hopping processes can stabilize the hexagon-shaped CLS.
In network superstructures, such destructive interference is expected to occur via the hopping processes at the lattice sites linking the neighboring 1D chains.
In the following, we explain how this is possible in network superstructures.

Examples of the network superstructures are illustrated in FIG.~1(a), (e), and FIG.~2(a).
First, the lattice structures in FIG.~1 are called network superstructures with dot-type links because the 1D chains are linked by purple-colored dots.
The cyclicgraphyne and cyclicgraphdiyne belong to this category.
Here, the 1D chain indicates the black dots in a straight line between two neighboring dot-type links.
The sites in a 1D chain are labeled as the $n$-th dot in the $m$-th chain.
The length of the $m$-th chain is represented by $Q_m$, which counts the number of sites in the $m$-th chain.
If the length of all the chains is the same, it is simply described by $Q$.
The amplitude of the CLS at the $n$-th dot in the $m$-th chain is denoted by $a_{m,n}$.
On the other hand, the amplitude of the CLS vanishes at the links.
The CLS can be stabilized in these lattice structures because the amplitudes at the neighboring sites of the dot-type link show destructive interference at the dot-type link after the hopping processes.

Second, the network structures in FIG.~2 are called network superstructures with triangle-type links because the 1D chains are linked by purple-colored triangular bonds.
The charge-density-wave phase of 1T-TaS$_2$ belongs to this class.
In this lattice structure, the amplitudes at the end sites of the neighboring 1D chains experience destructive interference via the triangular hopping structure.
For example, the amplitudes $a_{1,Q-1}$ and $a_{2,0}$ of the first and second 1D chains will meet at another site in the triangle where they belong to and vanish after the hopping processes if $a_{1,Q-1}=-a_{2,0}$.

\section{Tight-binding analysis}

\subsection{General recursion relation}\label{sec:gen_rec_rel}
The typical form of the CLSs of the network superstructures hosting flat bands is given by
\begin{align}
    |\chi \rangle = \sum_{m=1}^M \sum_{n=1}^{Q_m} a_{m,n}|m,n\rangle,
\end{align}
where $a_{m,n}$ is the amplitude of the CLS at the $n$-th atom of the $m$-th chain.
For the CLS to be an eigenstate of the given tight-binding Hamiltonian with the nearest-neighbor hopping parameter 1, the amplitudes should satisfy
\begin{align}
    a_{m,n} + a_{m,n-2} = \varepsilon a_{m,n-1},\label{eq:recursion_0}
\end{align}
where $\varepsilon$ is the energy of the flat band.
Without loss of generality, one can assume that $a_{m,0}=1$ releasing the normalization condition for the CLS.
Boundary conditions for the chain at two ends of it determine $a_{m,1}$ and $\varepsilon$.
Let us assume that we know the value of $a_{m,1}$ from one of the boundary conditions.
Then, the recursion relation (\ref{eq:recursion_0}) can be rewritten as
\begin{align}
    b_{m,n} = (\varepsilon - \alpha_\varepsilon) b_{m,n-1},\label{eq:recursion_1}
\end{align}
where $b_{m,n} = a_{m,n} - \alpha_\varepsilon a_{m,n-1}$. 
Here, $\alpha_\varepsilon$ satisfies $\alpha_\varepsilon = 1/(\varepsilon - \alpha_\varepsilon)$, namely,
\begin{align}
    \alpha_\varepsilon = \frac{\varepsilon \pm \sqrt{\varepsilon^2-4}}{2}.
\end{align}
From (\ref{eq:recursion_1}), we obtain
\begin{align}
    b_{m,n} = (\varepsilon - \alpha_\varepsilon)^{n-1}b_{m,1},
\end{align}
which leads to
\begin{align}
    a_{m,n} = F_{n+1}(\alpha_\varepsilon) + (a_{m,1}-\varepsilon)F_{n}(\alpha_\varepsilon),\label{eq:a_mn}
\end{align}
where
\begin{align}
    F_{n}(\alpha_\varepsilon) = \frac{\alpha_\varepsilon^{n}-\alpha_\varepsilon^{-n}}{\alpha_\varepsilon - \alpha_\varepsilon^{-1}}.\label{eq:F_function_0}
\end{align}
Since the eigenenergy of an infinite simple chain with the nearest-neighbor hopping parameter 1 lies between -2 and 2, it is reasonable to seek flat band energies in the same interval. 
Therefore, we can express the eigenenergy of the finite chain as $\varepsilon = 2\cos\theta$, which leads to 
\begin{align}
    \alpha_\varepsilon = e^{\pm i\theta }. 
\end{align}
Note that $\theta \neq 0$ to ensure the function $F_n$ is well-defined.
Then, $F_{n}(\alpha_\varepsilon)$ is simplified as
\begin{align}
    F_{n}(\alpha_\varepsilon) = \frac{\sin n\theta}{\sin \theta}.\label{eq:F_function_1}
\end{align}
The explicit forms of the amplitudes of the CLS and the corresponding eigenenergy are determined from the boundary conditions of the chain at the links attached to it.
The detailed forms of the eigenenergies of the flat bands of various network superstructures are derived in the following subsections.

\begin{figure}[ht]
    \centering
    \includegraphics[width=1.0\columnwidth]{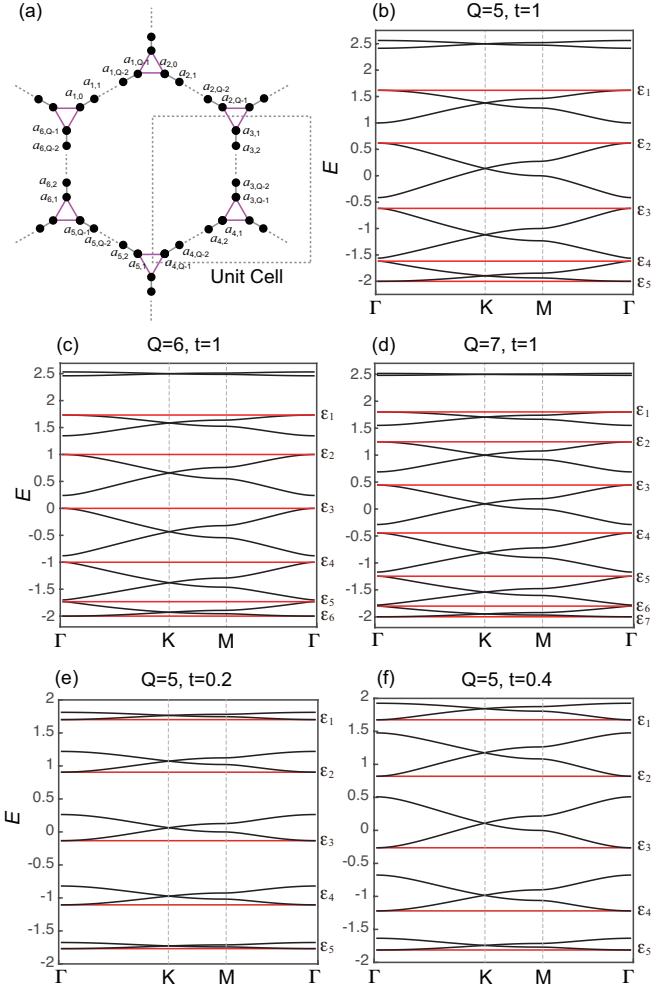}
    \caption{(a) The honeycomb network superstructure with the triangle-type links. The triangle-type links are colored purple. In the unit cell, there are $2Q$ sites. The hopping parameter for the triangle-type link is denoted by $t$, while it is 1 otherwise. From (b) to (d), we plot band spectra with $t=1$, while we consider $t\neq 1$ cases in (e) and (f). Flat bands are represented by red curves.}
    \label{fig:triangular_basic}
\end{figure}

\subsection{Network superstructures with dot-type links}
We first consider a square network superstructure, which contains two simple chains of the same size in a unit cell as illustrated in FIG.~1(a) with $Q_1=Q_2$.
The simple chains are connected to each other by dot-type links.
As shown in FIG.~1(a), we consider a CLS having zero amplitudes at the dot-type links.
As a result, for the chain with $m=1$, as an example, the recursion relation (\ref{eq:recursion_0}) for $n=1$ is given by
\begin{align}
    a_{1,1} = \varepsilon a_{1,0}=\varepsilon,
\end{align}
because $a_{m,n}$ is assumed to be zero for a negative $n$ and 1 for $n=0$.
By plugging-in this value of $a_{1,1}$ into (\ref{eq:a_mn}), the general term of the amplitude of the CLS along the $m$-th chain is obtained as
\begin{align}
    a_{1,n} = F_{n+1}(\alpha_\varepsilon),\label{eq:a_1n}
\end{align}
for $0\leq n \leq Q_1-1$.
\begin{figure}[ht]
    \centering
    \includegraphics[width=1.0\columnwidth]{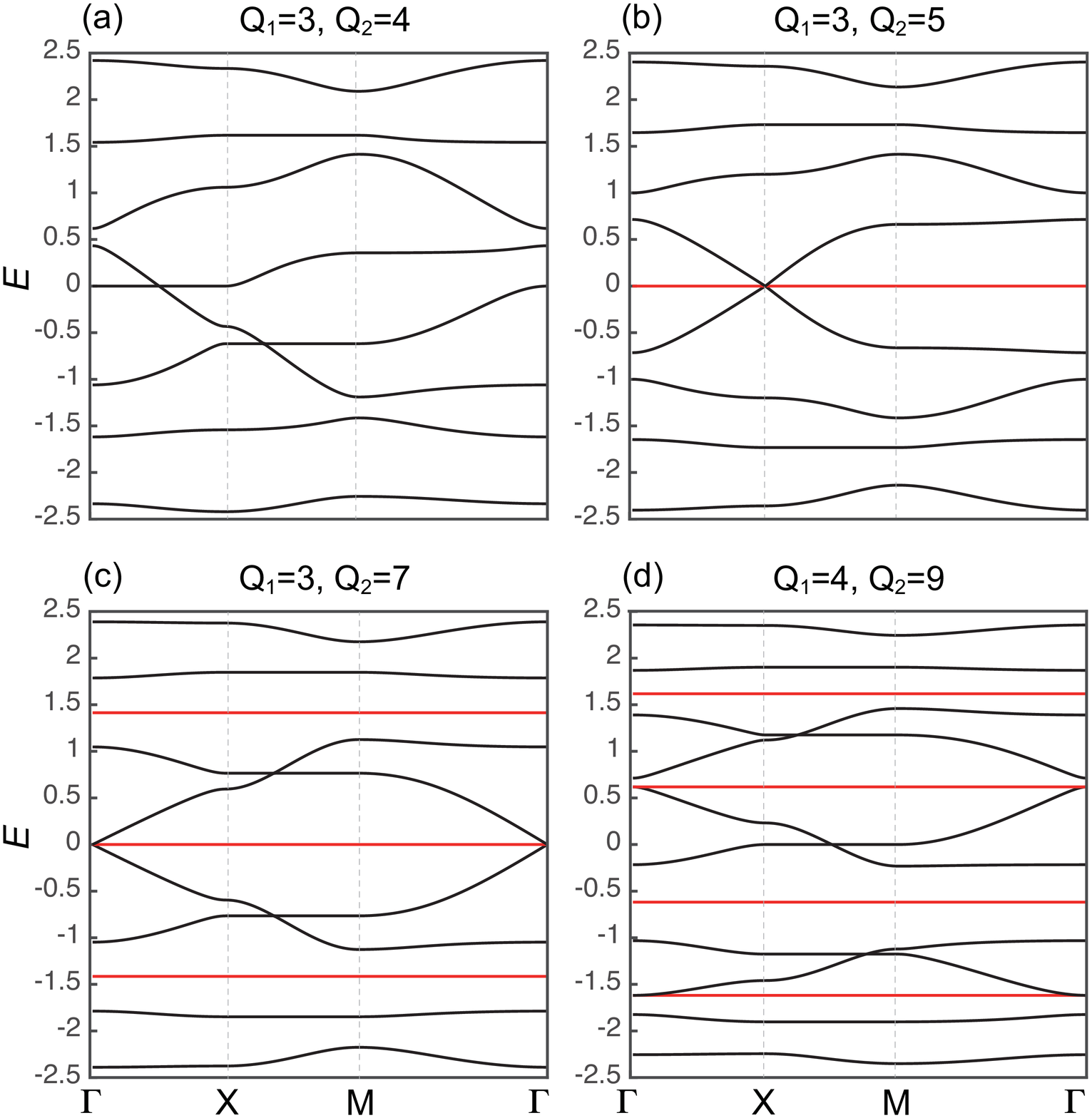}
    \caption{(a) Band spectra of rectangular network superstructures with $Q_1\neq Q_2$. Red lines denote flat bands. Refer to FIG.~1(a) for the lattice structure.}
    \label{fig:rectangular_nn}
\end{figure}
When $n=Q_1-1$, namely at the end of the simple chain, the recursion relation (\ref{eq:recursion_0}) becomes
\begin{align}
    a_{1,Q_1-2} = \varepsilon a_{1,Q_1-1},
\end{align}
because $a_{1,Q_1}$ is supposed to be zero.
By applying (\ref{eq:a_1n}) and $\varepsilon = 2\cos\theta$ to the above, we obtain
\begin{align}
   2\cos\theta \sin Q_1\theta = \sin (Q_1-1)\theta,
\end{align}
which can be simplified into
\begin{align}
    \sin (Q_1+1)\theta = 0.\label{eq:square_theta_eq}
\end{align}
From the solution of (\ref{eq:square_theta_eq}), $\theta = q\pi/(Q_1+1)$, where $q=1,~2,\cdots,~Q_1$, the eigenenergy of the flat band is evaluated as
\begin{align}
    \varepsilon_q = 2\cos \frac{q}{Q_1+1}\pi.\label{eq:fb_energy_1}
\end{align}
Note that $q=0$ and $q=Q_1+1$ are excluded because the corresponding eigenenergy is 2 which makes $\alpha_\varepsilon=\pm 1$ and $F_{n}$ ill-defined.
Also, we only considered positive $\theta$'s because $F_n(\alpha_\varepsilon)$ is an even function of $\theta$, and we obtain the same CLS and eigenenergy for $\theta$ and $-\theta$.
By noting that $a_{1,Q_1-1} = F_{Q_1}(\alpha_{\varepsilon_q})=(-1)^{q-1}$ for the flat band at $\varepsilon=\varepsilon_q$, the amplitudes of the CLS on other simple chains are given by $a_{2,n} = (-1)^q a_{1,n}$, $a_{3,n} = a_{1,n}$, $a_{4,n} = (-1)^q a_{1,n}$.
Since all the simple chains have the same size, $n$ runs from 0 to $Q_1-1$ for all $m$.
We confirmed that the formula (\ref{eq:fb_energy_1}) matches well with the flat band energies obtained numerically as shown in FIG.~1(b) to (d) for various values of $Q$.

The results of the eigenenergies and eigenfunctions of the CLSs in the above can be applied to any kind of network superstructures with dot-type links if the simple chains consisting of the network are of the same size.
For example, the CLS of the honeycomb network superstructure is illustrated in FIG.~1(e), and the destructive interference at the dot-type links can be hosted by letting the amplitudes of the CLS satisfy $a_{m,n}=(-1)^{m-1}a_{1,n}$.
The eigenenergies of the flat bands are completely described by (\ref{eq:fb_energy_1}).
The tight-binding calculations of the band structures of the honeycomb network of superstructures are exhibited in FIG.~1(f) to (h).

One can have flat bands even if the network superstructure consists of simple chains of different sizes($Q_1\neq Q_2$).
Refer to the rectangular network superstructure illustrated in FIG.~1(a).
The allowed eigenenergies of determined from these two chains are denoted by $\varepsilon_q^{(1)}=2\cos q\pi/(Q_1+1)$ and $\varepsilon_q^{(2)}=2\cos q\pi/(Q_2+1)$.
While the condition for the existence of a flat band is that the two chains share the same eigenenergy, there exist integers $q_1$ and $q_2$ that make $\varepsilon_{q_1}^{(1)} = \varepsilon_{q_2}^{(2)}$ if $Q_1+1$ and $Q_2+1$ have a common divisor larger than 1 and smaller than $\mathrm{min}(Q_1+1,Q_2+1)$.
In FIG.~3, we plot various examples of rectangular network superstructures with dot-type links.
As shown in FIG.~3(a), one cannot have a flat band because $Q_1+1=4$ and $Q_2+1=5$ have no common divisor.
On the other hand, if $Q_1=3$ and $Q_2=5$, one can have a flat band when $q_1=2$ and $q_2=3$ as illustrated in FIG.~3(b).
If $Q_2+1$ is an integer multiple of $Q_1+1$, we have $Q_1$ number of flat bands as plotted in FIG.~3(c) and (d).

\begin{figure*}[ht]
    \centering
    \includegraphics[width=2\columnwidth]{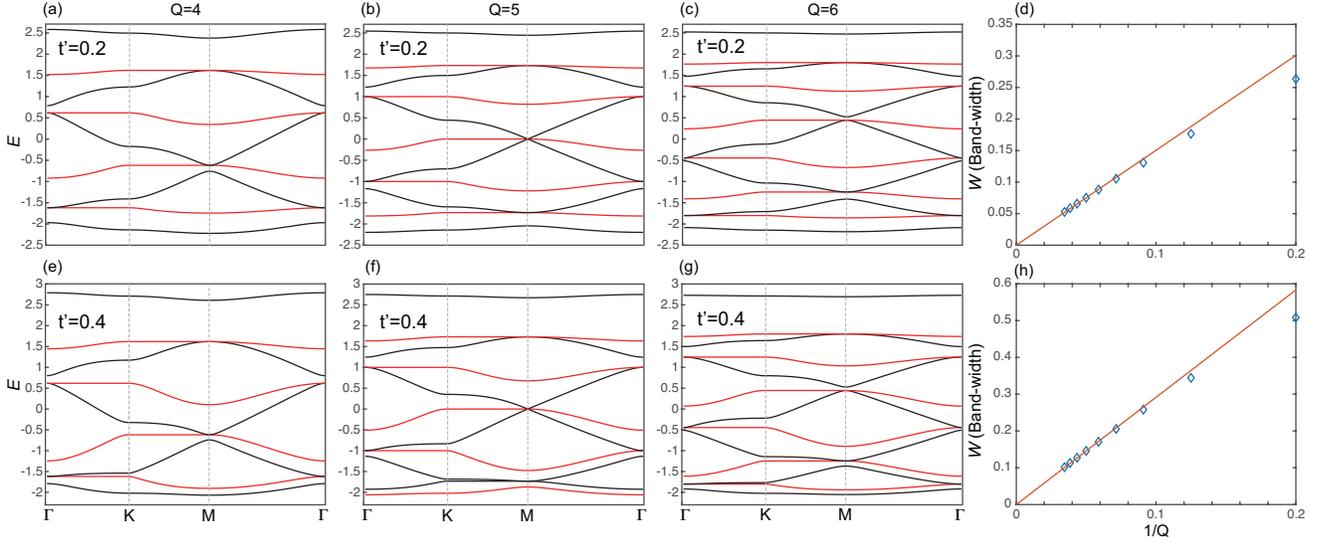}
    \caption{Band structures of the square network superstructures with the next nearest-neighbor hopping processes($t^\prime$) between four sites around the dot-type link in FIG.~1(a). The flat bands in FIG.~1(b) to (d) deform to the nearly flat bands colored red in (a) to (c) and (e) to (g) for $t^\prime=0.2$ and 0.4, respectively. In (d) and (h), we plot the bandwidth of the nearly flat band as a function of $1/Q$. Here, we take the maximum bandwidth among all the nearly flat bands for a given $Q$.}
    \label{fig:square_nnn}
\end{figure*}

\subsection{Network superstructures with triangle-type links}
The triangle-type links are illustrated in FIG.~2(a).
Unlike dot-type links, triangle-type links let the simple chains connect to each other directly.
Let us consider the honeycomb network consisting of simple chains of the same size $Q$.
We solve the recursion equation (\ref{eq:recursion_0}) for the $m=1$ simple chain, which neighbors with the $m=2$ and $m=6$ chains.
Assuming $a_{1,0}=1$, one can note that we should have $a_{6,Q-1}=-1$ to prevent the dissipation of the CLS via the destructive interference at the triangle-type link.
Then, the recursion relation relevant to the amplitudes of the CLS at the link sites is given by
\begin{align}
    a_{1,1} + a_{6,Q-1} = a_{1,1} -1 = \varepsilon a_{1,0} = \varepsilon,
\end{align}
which gives $a_{1,1} = \varepsilon +1$.
From (\ref{eq:a_mn}), the general term of the amplitude of the CLS on the $m=1$ simple chain is obtain as
\begin{align}
a_{1,n} &= F_{n+1}(\alpha_\varepsilon) + F_{n}(\alpha_\varepsilon),\\
&= \frac{\sin (n+1)\theta + \sin n\theta}{\sin\theta}.\label{eq:triangular_gen_term}
\end{align}
Due to the destructive interference at the other triangle link involving $a_{1,Q-1}$ and $a_{2,0}$, we have $a_{2,0}=-a_{1,Q-1}$, which results in $a_{2,n}=-a_{1,Q-1}[F_{n+1}(\alpha_\varepsilon) + F_{n}(\alpha_\varepsilon)] = -a_{1,Q-1}a_{1,n}$.
This means that $a_{2,Q-1} = -a_{1,Q-1}^2$.
By applying the same process to the other chains, we obtain $a_{6,Q-1} = -a_{1,Q-1}^6 = -1$, which implies that $a_{1,Q-1}=\pm1$ because the function $F_n(\alpha_\varepsilon)$ is real-valued.
Then, the eigenvalue condition is given by
\begin{align}
    \sin Q\theta + \sin (Q-1)\theta = \pm\sin\theta,
\end{align}
which leads to $\sin Q\theta/2 \times \cos (Q-1)\theta/2 =0$ or $\cos Q\theta/2 \times \sin (Q-1)\theta/2 =0$.
On the other hand, the eigenvalue equation at another boundary of the $m=1$ chain is given by
\begin{align}
    a_{1,Q-2} + a_{2,0} = a_{1,Q-2} - a_{1,Q-1} = \varepsilon a_{1,Q-1}.
\end{align}
Here, $a_{2,0}=a_{1,Q-1}$ to have destructive interference at the triangular link.
By using the identity $\varepsilon+1 = a_{1,1}$ and the formula (\ref{eq:triangular_gen_term}), we have another condition for $\theta$ given by $\sin Q\theta =0$.
These conditions for $\theta$ lead to the solutions of the form $\theta = \pi q/Q$, where $q$ is an integer running from 1 to $Q$.
Here, $q=0$ is excluded because this makes $F_n(\alpha_\varepsilon)$ ill-defined as explained in Sec.~\ref{sec:gen_rec_rel}.
These solutions of $\theta$ lead to the flat band energies given by
\begin{align}
    \varepsilon_q = 2\cos\frac{\pi q}{Q},\label{eq:e1_q}
\end{align}
by putting these $\theta$'s into $\varepsilon = 2\cos\theta$ derived in Sec.~\ref{sec:gen_rec_rel}.

%
%

\subsection{Triangle-type link with different hopping amplitudes}

The hopping parameters of the bonds in the triangle link can be different from 1, the hopping parameter of the simple chains, as shown in Fig.~.
Let us denote the hopping parameter in the link by $t$.
As in the previous cases, we set $a_{1,0}=1$.
Then, the recursion relation for determining $a_{1,1}$ is given by
\begin{align}
    a_{1,1}+t a_{6,Q-1} = a_{1,1} -t = \varepsilon a_{1,0} = \varepsilon,
\end{align}
which results in $a_{1,1} = \varepsilon +t$.
From, (\ref{eq:a_mn}), the general term is evaluated as
\begin{align}
    a_{1,n} &= F_{n+1}(\alpha_\varepsilon) + t F_{n}(\alpha_\varepsilon), \\
    &=\frac{\sin(n+1)\theta + t\sin n\theta}{\sin\theta}.
\end{align}
Eigenenergies of the flat bands are obtained from the condition $a_{1,Q-1}=\pm 1$, which leads to
\begin{align}
    \sin Q\theta + t\sin (Q-1)\theta \pm \sin\theta =0.\label{eq:triangular_tp_eq1}
\end{align}
Another boundary condition given by
\begin{align}
    a_{1,Q-2}+t a_{2,0} = a_{1,Q-2} - t a_{2,0} = \varepsilon a_{1,Q-1},
\end{align}
gives
\begin{align}
    \sin(Q-1)\theta + t \sin(Q-2)\theta \pm (\sin 2\theta + t\sin\theta) =0,\label{eq:triangular_tp_eq2}
\end{align}
where the sign $\pm$ is sinchronized with that of (\ref{eq:triangular_tp_eq1}).
By solving the coupled equations (\ref{eq:triangular_tp_eq1}) and (\ref{eq:triangular_tp_eq2}) with the same sign, we obtain the flat band energies $\varepsilon = 2\cos\theta$.

\subsection{Effect of the next-nearest-neighbor hopping processes}

Around the dot-type link in FIG.~1(a) and 1(e), let us consider the effect of the small next-nearest-neighbor hopping processes.
The corresponding parameter is denoted by $t^\prime$.
For example, in FIG.~1(a), the hopping processes between $(m,n)=(1,0)$ and $(m,n)=(4,Q-1)$, $(m,n)=(1,Q-1)$ and $(m,n)=(2,0)$, $(m,n)=(2,Q-1)$ and $(m,n)=(3,0)$, and $(m,n)=(3,Q-1)$ and $(m,n)=(4,0)$ are the next-nearest-neighbor ones.
As in many flat band models, the flat bands in the network superstructures are also easily warped when we include such long-range hopping interactions.
We obtain band dispersions of the square network superstructures with $t^\prime=0.2$ and $0.4$ as exhibited in FIG.~4.
The perfectly flat bands of the square network superstructures in FIG.~1 represented by red lines, deform to the nearly flat bands, also colored red, in FIG.~4.
We noted in FIG.~4(d) and (h) that the bandwidth of the nearly flat bands decreases as $1/Q$.
While the bandwidth of the nearly flat band scales with the effective hopping integral between localized modes at two neighboring 1D chains, the amplitudes at each site, such as $a_{1,0}$ and $a_{6,Q-1}$ are proportional to $1/\sqrt{Q}$.
As a result, the bandwidth is proportional to $t^\prime/Q$.
These results imply that one can flatten the nearly flat band as much as one wants by increasing the length of the 1D chains consisting of the network superstructures.

\section{Conclusions}
We have investigated the origin of the development of flat bands in network superstructures from the perspective of the stabilization of the CLSs because there is a correspondence between the flat band. 
Two types of the bonding structures linking 1D chains offer the destructive interference, so that the CLSs satisfy the eigenstate condition.
We have considered two types of links between 1D chains called dot-type and triangle-type links.
By using these links, we have constructed the square, rectangular, and honeycomb network superstructures and obtained analytic forms of the flat band energies and CLSs.
Note that the number of flat bands equals to the number of independent CLSs.
While the previously studied network superstructures such as cyclic-graphyne, cyclic-
graphdiyne, and honeycomb network in the nearly commensurate charge-density-wave phase of 1T-TaS$_2$ exhibit almost flat bands in spite of the long-range hopping processes, we demonstrated that this is because the overlap between localized modes in 1D chains scales as the inverse of the length of the chains.
Therefore, one can flatten the nearly flat band as much as we want by increasing the length of the 1D chains.

{\small \subsection*{Acknowledgements}
D.H, J.S.L, A.Z. and J.-W.R. were supported by the National Research
Foundation of Korea (NRF) Grant funded by the Korea government
(MSIT) (Grant No. 2021R1A2C1010572). J.-W.R. was supported by
the National Research Foundation of Korea (NRF) Grant funded by
the Korea government (MSIT) (Grant No. 2021R1A5A1032996).}


%

\end{document}